\documentclass[english,journal=jctcce,manuscript=article,email=true,etalmode=truncate,maxauthors=0,doi=true]{achemso}

\usepackage{graphicx}
\graphicspath{{figures/}}
\usepackage{physics}
\usepackage{multirow}
\usepackage{xcolor}
\usepackage{caption}
\usepackage{subcaption}
\usepackage{xcolor}
\usepackage{bm}
\usepackage{eufrak} 
\usepackage{yfonts} 
\usepackage[title,titletoc]{appendix}
\usepackage{diagbox}

\usepackage{hyperref} 
\hypersetup{colorlinks=true, citecolor=blue, urlcolor=blue, linkcolor=blue}
\usepackage{cleveref}
  	\crefname{figure}{Figure}{Figures}
  	\crefname{table}{Table}{Tables}
  	\crefname{equation}{Eq.}{Eqs.}
  	\crefname{section}{Section}{Sections}
  	\crefname{subsection}{Section}{Sections}
  	\crefname{subsubsection}{Section}{Sections}
  	\crefname{algorithm}{Algorithm}{Algorithms}

\newcommand{\doi}[1]{\href{http://dx.doi.org/#1}{\nolinkurl{#1}}}

\newcommand{\code}[1]{\texttt{#1}}
\newcommand\vartextvisiblespace[1][.5em]{%
  \makebox[#1]{%
    \kern.07em
    \vrule height.3ex
    \hrulefill
    \vrule height.3ex
    \kern.07em
  }
}
\usepackage{todonotes}

\usepackage{epstopdf}
\AppendGraphicsExtensions{.tiff}

\title{Direct determination of optimal real-space orbitals for correlated electronic structure of molecules.} 

\author{Edward F. Valeev}
\affiliation{Department of Chemistry, Virginia Tech, Blacksburg, VA 24061}
\email{efv@vt.edu}

\author{Robert J. Harrison}
\affiliation{Department of Applied Mathematics \& Statistics, Stony Brook University, Stony Brook, NY 11794}

\author{Adam A. Holmes}
\affiliation{Department of Chemistry, Virginia Tech, Blacksburg, VA 24061}

\author{Charles C. Peterson}
\affiliation{Office of Advanced Research Computing, University of California, Los Angeles, CA 90095}

\author{Deborah A. Penchoff}
\affiliation{UT Innovative Computing Laboratory, University of Tennessee, Knoxville, TN 37996}

\begin{document}

\date{\today}

\begin{abstract}
We demonstrate how to determine numerically nearly exact orthonormal orbitals that are optimal for evaluation of the energy of arbitrary (correlated) states of atoms and molecules by minimization of the energy Lagrangian. Orbitals are expressed in real space using a multiresolution spectral element basis that is refined adaptively to achieve the user-specified target precision while avoiding the ill-conditioning issues that plague AO basis set expansions traditionally used for correlated models of molecular electronic structure. For light atoms, the orbital solver, in conjunction with a variational electronic structure model [selected Configuration Interaction(CI)] provides energies of comparable precision to a state-of-the-art atomic CI solver. The computed electronic energies of atoms and molecules are significantly more accurate than the counterparts obtained with the Gaussian AO bases of the same rank, and can be determined even when linear dependence issues preclude the use of the AO bases. It is feasible to optimize more than 100 fully-correlated numerical orbitals on a single computer node, and significant room exists for additional improvement. These findings suggest that the real-space orbital representations might be the preferred alternative to AO representations for high-end models of correlated electronic states of molecules and materials.
\end{abstract}

\maketitle 

\section{Introduction}\label{sec:intro}

Predictive computation of properties of molecules and materials requires highly-{\em accurate} treatment of many-electron correlation effects as well as {\em precise} numerical representation of the many-body states and operators involved therein.
In molecular and, increasingly, in solid-state contexts many-body methods have utilized the Linear Combination of Atomic Orbitals (LCAO)\cite{VRG:roothaan:1951:RMP} numerical representation of 1-particle (orbital) states constructed from pre-optimized sequences of atom-centered basis sets (represented by analytic Gaussian-type\cite{VRG:boys:1950:PRSMPES} or Slater-type functions,\cite{VRG:zener:1930:PR,VRG:slater:1930:PR} or represented numerically\cite{VRG:blum:2009:CPC}).
However, (augmented) plane-wave (PW) representation of 1-particle states can also be used.\cite{VRG:schafer:2017:JCP,VRG:bylaska:2021:FC}
Unfortunately representation of many-particle by products of 1-particle states suffers from slow convergence in the vicinity of electron cusps,\cite{VRG:kato:1957:CPAM,VRG:pack:1966:JCP} thereby mandating the use of basis set extrapolation\cite{VRG:helgaker:1997:JCP} or
the augmentation with explicit 2-particle basis (``explicit correlation'')\cite{VRG:klopper:2006:IRPC,VRG:kong:2012:CR,VRG:hattig:2012:CR,VRG:gruneis:2013:JCP}.

Although the LCAO-based many-body methods can be pushed to match experimental thermochemical\cite{VRG:tajti:2004:JCP} and spectroscopic data\cite{VRG:barletta:2006:JCP} for small molecules, extending such successes to increasingly larger systems, and to higher accuracy, is limited by the rapid onset of ill-conditioning, high-order rise of operator evaluation with the AO angular momenta, and the challenges of fast (reduced-scaling) reformulation of such methods due to the increasing density of the representation in the high-precision regime. LCAO representation in practice is a heavily empirical endeavour: at this moment the Basis Set Exchange reference database\cite{VRG:pritchard:2019:JCIM} lists 402 Gaussian AO basis sets designed for representing orbitals of a carbon atom. Albeit the large number of officially-recognized bases is simply due to the long history of the Gaussian AO technology, many recently-designed bases have rather niche uses; navigating the basis set zoo is a near-expert task, especially when treating high-density solids\cite{VRG:vandevondele:2007:JCP,VRG:ye:2022:JCTC} or computing properties other than the energy.\cite{VRG:baranowska:2009:JCC} Meanwhile plane wave representation struggles with description of localized features (e.g., atomic shell structure, nuclear cusps and Coulomb holes) of the electronic wave functions and thus are not naturally suitable to compact and/or fast formulations.

An attractive alternative to the LCAO and PW representations for many-body electronic structure are {\em real-space} numerical representations, a loosely-defined group including finite difference (FD), which only use grids and lack explicit basis functions, and finite element (FE) and spectral element (SE) methods, which utilize basis functions with strictly local support (i.e., the basis functions are only nonzero in a finite domain). Such representations combine the ability to resolve short and long lengthscale features of the many-body electronic wave functions with promise of systematic bias-free improvability without ill-conditioning issues. Due to these favorable features there has been a flurry of activity\cite{VRG:chelikowsky:1994:PRL,VRG:briggs:1995:PRB,VRG:harrison:2004:JCP,VRG:fattebert:2006:PRB,VRG:genovese:2008:JCP,VRG:ohba:2012:CPC,VRG:motamarri:2013:JCP,VRG:yanai:2015:PCCP,VRG:jensen:2016:PCCP,VRG:xu:2019:JPCM} in developing real-space numerical technology for fast 1-body methods, such as Kohn-Sham density functional theory.

Unfortunately the use of real-space representation has been far more limited for many-body methods applicable to general molecules due to the exponential growth of the representation  of a $k$-particle state with $k$. Even methods like MP2 and CCSD in first-quantized formulation\cite{VRG:jeziorski:1984:JCP} involve representation of 2-particle wave operators (hence, 6-dimensional functions); while for atoms pair theories have been explored for a relatively long time,\cite{VRG:flores:1993:JCP,VRG:ackermann:1995:PRA,VRG:flores:1999:JPBAMOP,VRG:flores:2008:IJQC}
for molecules the feasibility of such representations has been demonstrated only recently by some of us (MP2)\cite{VRG:bischoff:2012:JCP,VRG:bischoff:2013:JCP} and our collaborators (CIS(D), CC2)\cite{VRG:kottmann:2017:JCTC,VRG:kottmann:2017:JCTCa} by combining volume-element-wise pair function compression and regularization of electron cusps\cite{VRG:bischoff:2012:JCP} similar to that performed in explicitly correlated R12/F12 methods.\cite{VRG:kong:2012:CR} It is possible to obtain energies and properties with such methods that are competitive to the state-of-the-art LCAO technology for small polyatomics, but the methodology is still relatively limited and likely cannot be made practical 3- and higher-body methods.

A more practical alternative to the direct real-space representation of 2-particle states is to expand them in products of real-space orbitals (also referred to as the algebraic approximation\cite{VRG:wilson:1976:PRA}). Although such approaches, unlike the direct many-body representations, suffer from the slow convergence near electron cusps, the same remedies as used for the LCAO-based many-body methods (namely, the basis set extrapolation or R12/F12-style explicit correlation) can be deployed to fix these shortcomings. In the purely orbital-based many-body framework the key differences between spectral (LCAO, PW) and real-space representations is limited to the orbital optimization and evaluation of the physical operators. Due to the cost of both in the real-space representation, real-space many-body methods have been implemented for atoms (both at the multiconfiguration self-consistent field (MCSCF)\cite{VRG:fischer:1968:A,VRG:fischer:1972:ADaNDT,VRG:sundholm:1989:NDotESoADaPM} and perturbative levels\cite{VRG:johnson:1986:PRL,VRG:dzuba:1996:PRA}) and diatomic molecules (MCSCF,\cite{VRG:adamowicz:1981:JCP,VRG:laaksonen:1984:CPL,VRG:sundholm:1989:NDotESoADaPM} perturbation theory\cite{VRG:mccullough:1984:FSCS} and coupled-cluster (CC)\cite{VRG:adamowicz:1985:JCP,VRG:adamowicz:1985:PRL}); an excellent recent review of the vast relevant literature can be found in Ref. \citenum{VRG:lehtola:2019:IJQC}.

The first demonstration of any real-space orbital-based many-body method applicable to general molecules was shown recently by Kottmann, Bischoff, and Valeev\cite{VRG:kottmann:2020:JCP}. The key to this advance was the realization that far more compact orbital representation of the weakly-occupied (virtual) orbitals needed for dynamical correlation is provided by the pair-natural orbitals.\cite{VRG:edmiston:1965:JCP,VRG:meyer:1971:IJQC,VRG:neese:2009:JCP} The PNO-MP2-F12 method using the real-space representation for all orbitals was demonstrated for molecules with more than 30 atoms, with precision surpassing what is achievable by the state-of-the-art LCAO technology.

While the work showed a viable roadmap for developing many-body perturbation theory (MBPT) and related methods with real-space orbitals, a single Slater determinant often does not provide a qualitatively correct starting point. In this paper we ask whether it is possible
\begin{enumerate}
\item to express arbitrary (not just MBPT-based) orbital-based many-body wave functions in terms of real-space orbitals, and
\item robustly determine optimal design for such orbitals.
\end{enumerate}
We conclude that the answer to these questions is affirmative.

Note that much the formalism for optimizing correlated real-space orbitals
presented in this work is generic and can be used with any real-space representation. The specific numerical representation that we used here
is based on the multiresolution analysis (MRA) of a Hilbert space that
leads to adaptive construction of scale-invariant real-space discountinuous high-order spectral element bases.\cite{VRG:alpert:1990:}
The resulting multiresolution adaptive spectral element representation (or, simply, MRA representation)\cite{VRG:beylkin:1991:CPAM,VRG:alpert:1993:SJMA,VRG:alpert:2002:JoCP,VRG:beylkin:2008:ACHA}
is usable for treating describing both valence and compact (core) electronic states with robust control of numerical error $\epsilon$. The multiscale structure of the representation results in low-order computational complexity for treating large systems. A number of electronic structure methods have been demonstrated in the multiresolution representation, including one-body (Hartree-Fock and DFT\cite{VRG:harrison:2004:JCP}) and 2-body (MP2\cite{VRG:bischoff:2012:JCP,VRG:bischoff:2013:JCP}, CC2\cite{VRG:kottmann:2017:JCTC}) methods, with energies, forces, and other molecular properties\cite{VRG:yanai:2004:JCP,VRG:kottmann:2017:JCTCa,VRG:bischoff:2017:JCP} of ground and excited states.
The complete technical details of the numerical representation can be found in Refs.~\cite{VRG:alpert:2002:JoCP,VRG:harrison:2004:JCP,VRG:harrison:2016:SJSC}; for a pedagogical introduction the reader is referred to Ref. \citenum{VRG:bischoff:2011:JCP}.

The rest of the paper is organized as follows. Section \ref{sec:formalism} describes the correlated multiresolution orbital solver. Section \ref{sec:technical} describes the technical details of the computational experiments that are reported and analyzed in Section \ref{sec:results}. Section \ref{sec:summary} summarizes our findings and outlines future directions.

\section{Formalism}\label{sec:formalism}
Consider the energy functional of a (unit-normalized) state $\Psi$ in a Fock-space state defined by $M$ (spin-)orbitals $\{\phi_p\}$:
\begin{align}
\label{eq:E}
    E(\ket{\Psi}) = \bra{\Psi} \hat{H} \ket{\Psi} = \gamma^j_i h^i_j + \frac{1}{2} \gamma_{ij}^{kl} g^{ij}_{kl},
\end{align}
where $\gamma^i_j$ and $\gamma^{ij}_{kl}$ are the (spin-orbital) 1- and 2-RDMs (see Appendix \ref{sec:notation} for the notation and standard definitions).
For a fixed {\em model} defining the map from orbitals $\{\phi_p\}$ to $\ket{\Psi}$, \cref{eq:E} defines a functional of the orbitals.
Our objective is to minimize this functional with respect to $M$ {\em orthonormal} orbitals $\{\phi_p\}$. Unlike the traditional representations of orbitals in terms of a fixed basis, such as an atomic orbital basis in the LCAO formalism, in this work each orbital is expressed in a basis {\em adaptively} refined to approach the orbital's exact representation in the full Hilbert space. The lack of a fixed basis precludes automatic enforcement of the orbital orthonormality by, e.g., unitary parametrization. Therefore it will be necessary to maintain orthonormality explicitly.

Here we follow the standard Lagrangian formalism for introducing orbital orthonormality constraints. Variation of the Lagrangian
\begin{align}
    L = \bra{\Psi}\hat{H}\ket{\Psi} - \epsilon^i_j (s_i^j - \delta_i^j),
\end{align}
by replacement
$\phi_m^* \to \phi_m^* + \delta$ produces the following first-order change:
\begin{align}
\label{eq:Lgrad0}
    \ket{\frac{\partial L}{\partial \phi_m^*}} = 2 \left( \gamma^m_i \hat{h} +
    \gamma^{mk}_{ij} \hat{g}^{j}_{k} - \epsilon^m_i \right) \ket{\phi_i} ,
\end{align}
with $\hat{g}^p_q$ defined in \cref{eq:hatop2idx}.
The optimal orbitals are determined by solving
\begin{align}
\label{eq:cond}
\ket{\frac{\partial L}{\partial \phi_m^*}} \overset{\text{opt}}{=} 0
\end{align}
The multipliers are determined by projecting \cref{eq:cond} onto the current (orthonormal) orbitals:
\begin{align}
    \bra{\phi_n}\ket{\frac{\partial L}{\partial \phi_m^*}} = & \, 2 \left( \gamma^m_i h^i_n +
    \gamma^{mk}_{ij} g^{ij}_{nk} - \epsilon^m_n \right) = 0, \quad \text{or} \\
    \label{eq:eps0}
    \epsilon^m_n = & \, \gamma^m_i h^i_n +
    \gamma^{mk}_{ij} g^{ij}_{nk}.
\end{align}
This choice of the multipliers makes the gradient automatically orthogonal to the current orbitals.

If $\ket{\Psi}$ is a single Slater determinant (SD) (e.g., a Hartree-Fock determinant) the RDMs simplify,
\begin{align}
\label{eq:gamma1-sd}
    \gamma^i_j \overset{\text{SD}}{=} & \, \delta^i_j,\\
    \gamma^{ij}_{kl} \overset{\text{SD}}{=} & \, \delta^i_k \delta^j_l - \delta^i_l \delta^j_k,
\end{align}
and the gradient
becomes:
\begin{align}
\label{eq:LgradSD}
    \ket{\frac{\partial L}{\partial \phi_m^*}} \overset{\text{SD}}{=} 2  \left( \hat{f} \ket{\phi_m} - f^m_i \ket{\phi_i} \right).
\end{align}
Its projection onto an arbitrary orbital $\phi_a$ orthogonal to all current orbitals is
\begin{align}
    \bra{\phi_a}\ket{\frac{\partial L}{\partial \phi_m^*}} \overset{\text{SD}}{=} 2 f^m_a ,
\end{align}
vanishing of which with the optimal orbitals is the Brillouin condition.

For multiconfigurational states \cref{eq:Lgrad0,eq:eps0} for the Lagrangian gradient and multipliers take the SD-like form when the 2-RDM is rewritten in terms of its cumulant (see Appendix \ref{sec:notation}). This nominally results in replacements $h\to f$, $\gamma\to\lambda$:
\begin{align}
    \label{eq:Lgrad1}
    \ket{\frac{\partial L}{\partial \phi_m^*}} 
    = & 2 \left( \gamma^m_i \hat{f}  +
    \lambda^{mk}_{ij} \hat{g}^{j}_{ k} - \epsilon^m_i \right) \ket{\phi_i} , \\
    \label{eq:eps1}
    \epsilon^m_n
    = & \, \gamma^m_i f^i_n +
    \lambda^{mk}_{ij} g^{ij}_{nk}.
\end{align}
Clearly, \cref{eq:LgradSD} follows from \cref{eq:Lgrad1} due to the vanishing 2-RDM cumulant for the SD state.

Note that the single determinant expression for the gradient with respect to orbital $\phi_m$ (\cref{eq:LgradSD}) only involves the action of the Fock operator onto $\phi_m$ itself.
In contrast, the $\gamma^m_i \hat{f} \ket{\phi_i}$ contribution to the multideterminantal expression \eqref{eq:Lgrad3}  involves the action of the Fock operator on {\em every} orbital.
The use of {\em natural}
orbitals (NOs), which make the 1-RDM diagonal,
\begin{align}
    \gamma^m_n \overset{\rm NO}{=} \begin{cases} \gamma^m_m, & \quad m=n \\
    0, & \quad m \neq n
    \end{cases}
\end{align}
allows to eliminate the {\em kinematic} coupling between the gradients:
\begin{align}
    \label{eq:Lgrad2}
    \ket{\frac{\partial L}{\partial \phi_m^*}} \overset{\text{NO}}{=} & 2 \left( \gamma^m_m \hat{f}\ket{\phi_m} + \left(
    \lambda^{mk}_{ij} \hat{g}^{j}_{k} - \epsilon^m_i \right) \ket{\phi_i} \right).
\end{align}

Until now the formalism has been completely generic, i.e., applicable to any real-space representation. However, the manner in which \cref{eq:cond} must be solved differs between numerical representation. Thus it is now necessary to specialize the formalism to the MRA numerical representation used in this work that we introduced in \cref{sec:intro}. Orbital optimization in the MRA representation in both SD\cite{VRG:harrison:2004:JCP} and correlated\cite{VRG:kottmann:2020:JCP} contexts uses the integral reformulation of the differential equations (this is crucial for being able to solve differential equations in {\em discontinuous} spectral element basis).
Partitioning the Fock operator $\hat{f}$
into the free-particle Hamiltonian $\hat{d}$ and potential $\hat{v}$ (see \cref{eq:f-dv}), the orbital stationarity condition becomes:
\begin{align}
    \label{eq:Lgrad3}
    - \left(\hat{d} - 
    \tilde{\epsilon}_m \right) \ket{\phi_m}
    = \hat{v}\ket{\phi_m} + \frac{1}{\gamma^m_m} \left(\gamma^{mk}_{ij} \hat{g}^{j}_{k} \ket{\phi_i} - \sum_{i\neq m} \epsilon^m_i \ket{\phi_i}\right),
\end{align}
where
\begin{align}
\tilde{\epsilon}_m \equiv \frac{\epsilon^m_m}{\gamma^m_m} \overset{\mathrm{NO}}{=} f^m_m + \sum_{ijk} \frac{\lambda^{mk}_{ij} g^{ij}_{mk}}{ \gamma^m_m} = h^m_m + \sum_{ijk} \frac{\gamma^{mk}_{ij} g^{ij}_{mk}}{ \gamma^m_m}
\end{align}
is the effective energy of NO $m$.
Note that the natural orbital basis is essential for the integral reformulation since it eliminates all but a single instance of the differential operator $\hat{d}$; solving for the natural orbitals is thus most sound strategy in the MRA numerical representation!

The orbital update described below exploits the fact that $\tilde{\epsilon}_m < 0$ for all orbitals, even weakly occupied. This may appear counterintuitive, since we normally think of unoccupied orbitals as corresponding to positive energies. To rationalize this observation, consider decomposition of the electronic energy, \cref{eq:E}, into the contributions from natural orbital $m$ and the rest:
\begin{align}
  E \equiv& E_m + E_{-m} \nonumber \\
  E_m \equiv& 2 \Re \left( \gamma_m^m h^m_m +  \sum_{ijk} \gamma_{ij}^{km} g^{ij}_{km}  \right) = 2 \Re \epsilon_m^m.
\end{align}
Since the electronic energy becomes more negative as the orbital basis is expanded, $E_m$ approximates the absolute {\em decrease} in the energy due to the addition of orbital $m$ to the orbital set. A more rigorous mathematical analysis may be warranted, but the simple argument should suffice for now.

Updated orbitals are obtained by the action of
Green's operator of particle in constant potential
$-\tilde{\epsilon}_m > 0$ on the r.h.s. of \cref{eq:Lgrad3}:
\begin{align}
\label{eq:phi-update}
    \ket{\phi_m} = & - \hat{G}_{-\tilde{\epsilon}_m} \left(\hat{v}\ket{\phi_m} + \frac{1}{\gamma^m_m} \left(\lambda^{mk}_{ij} \hat{g}^{j}_{k} \ket{\phi_i} - \sum_{i\neq m} \epsilon^m_i \ket{\phi_i}\right)\right).
\end{align}
For a nonrelativistic particle and open boundary conditions Green's operator has the familiar Yukawa kernel:
\begin{align}
    x>0: \hat{G}_{-x} f(\mathbf{r}) = - 2 \int \frac{\exp(-\sqrt{2x}|\mathbf{r}-\mathbf{r}'|)}{4 \pi |\mathbf{r}-\mathbf{r}'|} f(\mathrm{r}') \, \mathrm{d}\mathbf{r}'
\end{align}
extensions to the periodic boundary conditions\cite{VRG:jia:2009:P8ISDCABES} and Dirac particles\cite{VRG:anderson:2019:JCP} are also known.
The updated orbitals are not orthonormal, thus the standard symmetric (L\"owdin) orthonormalization\cite{VRG:lowdin:1950:JCP} is applied to restore the orbital orthonormality while minimizing the induced change to facilitate extrapolation.\cite{VRG:mayer:2002:IJQC}

For highly-symmetric cases, like NOs with high angular momentum in atoms, it may be also desirable to project orbitals onto the corresponding invariant subspaces of the symmetry group. This is due to the spectral element basis lacking the desired symmetry properties, thus symmetries of the orbitals are maintained to precision $\epsilon$ used to define the basis; this is further compounded by the additional approximations invoked when evaluating \cref{eq:phi-update}.
Appendix \ref{sec:spherical-projection} described the relevant projection scheme used for the case of atoms where the problem can become particularly severe when low precision ($\epsilon>10^{-4}$) is used.

The rate of convergence of \cref{eq:phi-update} is rapid, but it is possible to accelerate it further by extrapolation. To this end we deploy the Krylov-subspace Accelerated Inexact Newton (KAIN) method;\cite{VRG:harrison:2004:JCC} other strategies like the Broyden–Fletcher–Goldfarb–Shanno (BFGS) method\cite{VRG:yao:2021:JCTC} should be usable as well. Note that to be able to make KAIN extrapolation robust it is essential to choose a canonical frame for the orbitals.
Detailed description of the canonicalization approach are outside of the scope of this article (canonicalization is only essential for accelerating the solver) and will be reported elsewhere.

It is useful to point out how \cref{eq:phi-update} is related to the orbital update used to determine optimal single-determinant (SD) orbitals in Ref. \citenum{VRG:harrison:2004:JCP}.
For a single determinant the cumulant vanishes,
\begin{align}
\epsilon^m_n \overset{\mathrm{SD}}{=} & f^m_n, \quad \text{and}\\
\gamma^m_n \overset{\mathrm{SD}}{=} & \delta^m_n;
\end{align}
hence the orbitals are automatically natural and the gradients \eqref{eq:LgradSD} are kinematically decoupled.
The residual coupling in the SD version of \eqref{eq:phi-update} is via the Lagrange multipliers. Such coupling
can be removed completely by switching to the {\em canonical} orbitals which make
the Fock operator diagonal ($f^m_n = \delta^m_n \epsilon_m$) and simplify the orbital update to
\begin{align}
    \ket{\phi_m} \overset{\mathrm{SD}}= & - \hat{G}_{-\epsilon_m} \hat{v}\ket{\phi_m}.
\end{align}
This coincides with the standard orbital update used for single determinant SCF solver in \code{MADNESS} when orbitals are not localized; with localized orbitals
the SD version of \cref{eq:phi-update} (with all cumulant contributions omitted) is used instead.
Note that in all of these approaches the Lagrange multipliers are evaluated every iteration directly (as matrix elements of the Fock operator in the basis of current orbitals) rather than iteratively updated
as in the original solver described in Ref. \citenum{VRG:harrison:2004:JCP}.

Specialization of \cref{eq:eps1,eq:phi-update} for the important case of the complete active space is straightforward, namely for inactive orbitals the cumulant contributions vanish. No additional changes are needed.

\section{Technical Details}\label{sec:technical}

The orbital solver has been implemented in a developmental version of the Massively Parallel Quantum Chemistry (\code{MPQC}) package\cite{VRG:peng:2020:JCP} using the implementation of the MRA calculus in the \code{MADNESS} library.\cite{VRG:harrison:2016:SJSC} The solver is bootstrapped with Gaussian AOs taken from the user-provided basis and projected onto the spectral element basis. Mean-field orbitals are then obtained by optimization within the projected AO basis. RDMs were obtained from 1- and 2-electron integrals evaluated in the mean-field (SCF) orbitals expressed in real-space basis using a Heat-Bath CI (HCI)\cite{VRG:holmes:2016:JCTC} solver developed in \code{MPQC}. All computations on atoms and the minimal-basis H$_{10}$ computation set HCI selection thresholds $\epsilon_{1,2}$ to zero; the H$_{10}$ computation bootstrapped from the cc-pV\{D,T\}Z bases used $\epsilon_{\{1,2\}}=10^{\{-4,-6\}}$.

\section{Results}\label{sec:results}

\subsection{Atoms}
\subsubsection{He}

The CI energies obtained with fixed Gaussian AO bases and the corresponding spectral-element orbitals obtained therefrom are shown in \cref{tab:he-energies}.

Unsurprisingly, the use of optimized numerical orbitals results in significant lowerings of the CI energy. Clearly, orbital optimization alone only addresses the suboptimality of the given orbital set of a fixed finite rank (``orbital error'', quantified by $E_{\rm AO} - E_{\rm MRA}$); the basis set incompleteness due to the finite rank (``rank error'', quantified by $E_{\rm MRA} - E_{\rm exact}$) is not addressed by the orbital optimization and requires increasing the rank (ideally, in conjunction with extrapolation or explicit correlation).
Nevertheless, it is somewhat surprising how substantial is the the orbital errors of the correlation consistent basis sets (that are optimized for atomic CI energies) relative to their rank errors, shown in the second-to-last column in \cref{tab:he-energies}. Even for a quadruple-zeta basis orbital relaxation lowers the energy by $~60\%$ of the residual basis set error. These findings are in line with the unexpected sensitivity of the CC singles and doubles energy to the details of the unoccupied (pair-natural) orbitals that we found in the LCAO context.\cite{VRG:clement:2018:JCTC}

What is more surprising, however, is that usually significant ($>50\%$) fraction of the energy lowering due to the orbital optimization can be attributed to the relaxation of the weakly-occupied (correlating) orbitals. To estimate this fraction the last column shows the the energy lowering due to relaxing the occupied Hartree-Fock (1s) orbital from its finite basis form to the exact form, as a percentage of the total CI energy lowering due to the orbital optimization. For example, with the cc-pVTZ AO basis only 33 \% of the CI energy lowering can be attributed to the basis set incompleteness of the Hartree-Fock occupied orbital; this is somewhat unexpected considering that the correlation consistent basis family was optimized for atomic CI computations. This finding emphasizes that even numerical deficiencies of
even weakly occupied orbitals can contribute significantly to the overall basis set errors of correlated energies.

\begin{table}[!ht]
    \centering
    \begin{footnotesize}
    \begin{tabular}{l|rr|rr|rr}
    \hline\hline
       Basis$^a$ & \multicolumn{2}{c}{AO} & \multicolumn{2}{c}{MRA} & \multicolumn{2}{c}{AO vs MRA} \\
        & $E_{\rm AO}$ & $E_{\rm AO}-E_{\rm exact}$ & $E_{\rm MRA}$ & $E_{\rm MRA}-E_{\rm exact}$ & $\frac{E_{\rm AO} - E_{\rm MRA}}{E_{\rm MRA} - E_{\rm exact}} \times 100\%$ & $\frac{E^{\rm HF}_{\rm AO} - E^{\rm HF}_{\rm exact}}{E_{\rm AO} - E_{\rm MRA}} \times 100\%$ \\
        \hline
1s & -2 846.292 & 57.43 & -2 861.679 & 42.05 & 37 & 100 \\
2s1p & -2 887.595 & 16.13 & -2 897.674 & 6.05 & 167 & 64 \\
3s2p1d & -2 900.232	& 3.49 & -2 901.840 & 1.88 & 86 & 33 \\
4s3p2d1f & -2 902.411 &	1.31 & -2 902.909 & 0.82 & 61 & 34 \\
5s4p3d2f1g & -2 903.152 & 0.57 & -2 903.297 & 0.43 & 35 & 40 \\
    \hline\hline
    \end{tabular}
    
    $^a$ Gaussian computations used the standard STO-6G\cite{VRG:hehre:1969:JCP,VRG:hehre:1970:JCP} and cc-pV\{D,T,Q,5\}Z\cite{VRG:woon:1994:JCP} basis sets, respectively. MRA orbitals were obtained by optimization starting from the Gaussian AO basis sets. 
    \end{footnotesize}
    \caption{Converged CI energies ($mE_{\rm h}$) of the ground-state He atom evaluated with fixed Gaussian AO and optimized MRA orbitals. The exact energy is $-2903.724\,mE_{\rm h}$. The energies are converged to all digits shown.}
    \label{tab:he-energies}
\end{table}

\begin{figure}[!ht]
\begin{tabular}{ccc}
  \includegraphics[width=0.33\textwidth]{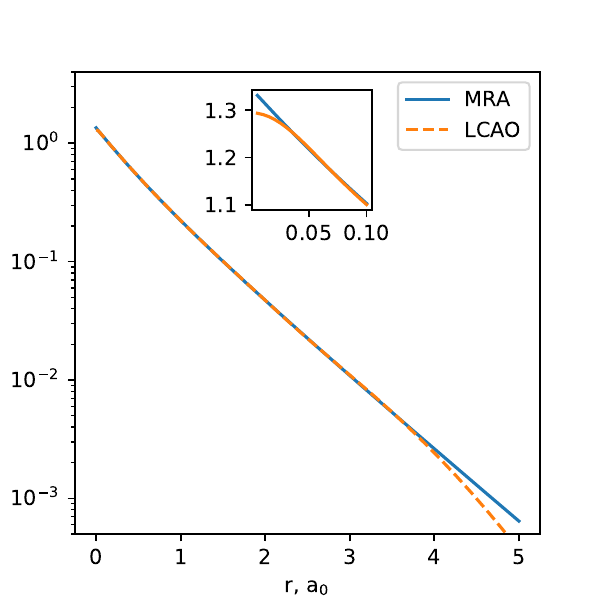} &
  \includegraphics[width=0.33\textwidth]{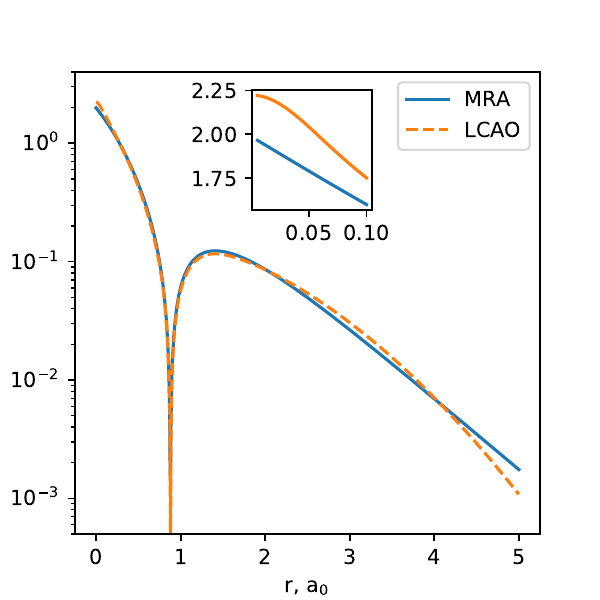} &
  \includegraphics[width=0.33\textwidth]{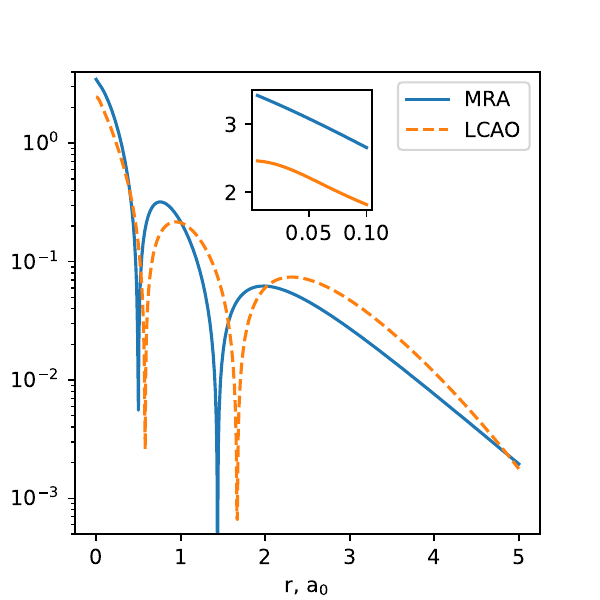}
\\
(a) 1st $l=0$ NO & (b) 2nd $l=0$ NO & (c) 3rd $l=0$ NO\\
  \includegraphics[width=0.33\textwidth]{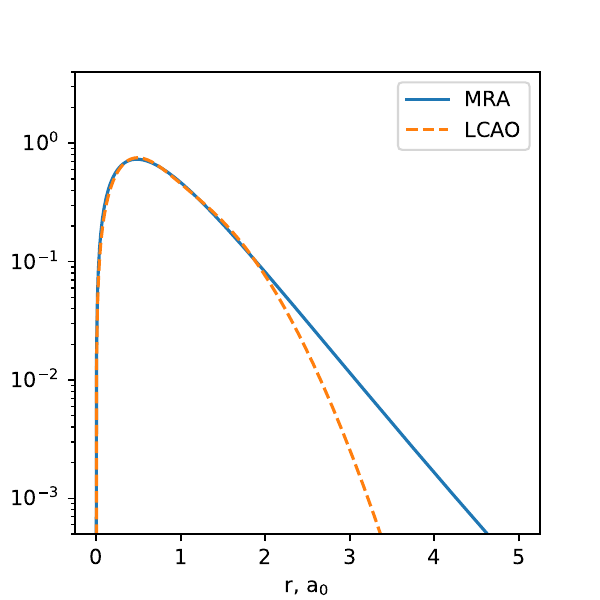} &
  \includegraphics[width=0.33\textwidth]{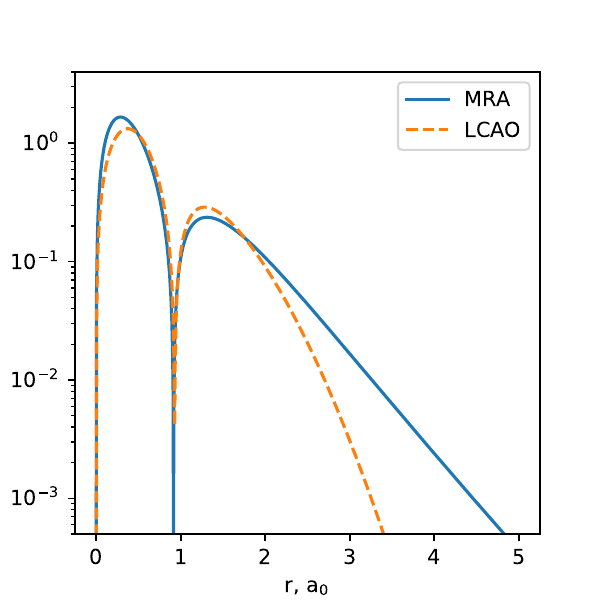}
\\
(d) 1st $l=1$ NO & (e) 2nd $l=1$ NO \\
  \includegraphics[width=0.33
\textwidth]{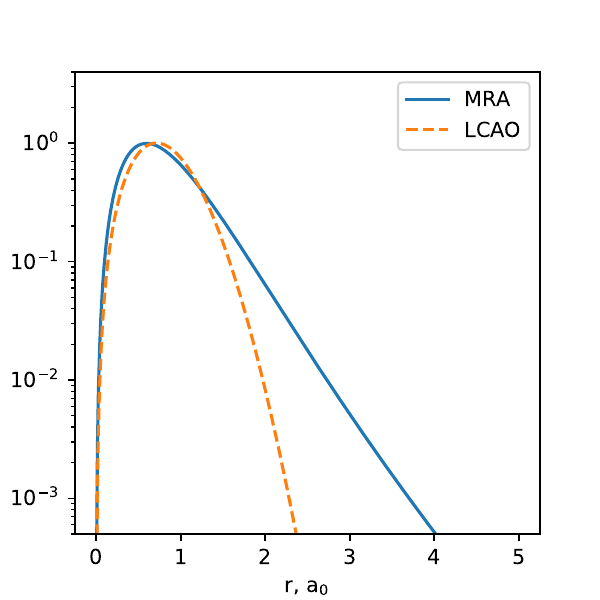}
\\
(f) 1st $l=2$ NO \\
\end{tabular}
\caption{Comparison of 6 largest-occupancy NOs of the ground-state He atom evaluated using Gaussian cc-pVTZ basis (``LCAO'') and the real-space orbitals (``MRA'') obtained therefrom. Gaussian-based NOs decay too rapidly in the asymptotic limit as well as fail to reproduce the cusp at the nucleus for the $l=0$ case.}
\label{fig:he-orbitals}
\end{figure}

As \cref{fig:he-orbitals} illustrates, the changes in the natural orbitals due to the basis set relaxation can indeed be significant, not only near the nucleus (where Gaussian orbitals lack the cusp) and far from the nucleus (where Gaussians decay too quickly) but also at the intermediate distances from the nucleus where the orbital magnitude and positions of the radial nodes can be strongly impacted by the orbital optimization.

As illustrated in \cref{tab:he-orbital-energies}, the orbital solver can robustly optimize orbital sets with occupancies differing by many orders of magnitude. As elaborated in \cref{sec:formalism}, the effective orbital energy, $\tilde{\epsilon}$, that controls the orbital update via the Green's operator (\cref{eq:phi-update}) is negative even for the weakly-occupied NOs; in fact, they are even more negative that the energy of the strongly occupied NO. In contrast, the corresponding diagonal elements of the Fock operator are positive, as expected. The difference between the Fock and effective orbital energies is negligible for the strongly occupied NO.

\begin{table}[h]
    \centering
    \begin{tabular}{c|lrr}
    \hline \hline 
        $\phi_p$ & \multicolumn{1}{c}{$\gamma^p_p$} & \multicolumn{1}{c}{$F^p_p$, $E_{\rm h}$} & \multicolumn{1}{c}{$\tilde{\epsilon}^p_p$ , $E_{\rm h}$} \\ \hline
1s	& 1.984	& -0.913& -0.958 \\
2s	& $7.69 \times 10^{-3}$	& 1.595	& -2.935 \\
2p	& $2.60 \times 10^{-3}$	& 2.058	& -3.305 \\
3s	& $1.08 \times 10^{-4}$	& 7.400	& -8.319 \\
3p	& $7.05 \times 10^{-5}$	& 7.879	& -8.789 \\
3d	& $5.85 \times 10^{-5}$	& 5.927	& -6.997 \\
\hline
    \end{tabular}
    \caption{Properties of the leading natural orbitals for the He ground state obtained from the cc-pVTZ basis.}
    \label{tab:he-orbital-energies}
\end{table}

\subsubsection{Be}

For the ground state of Be the energy lowering due to the orbital relaxation was similar to what was found for He (see Table \cref{tab:be-energies}. Microhartree-level agreement was found with the atomic CI solver in the state-of-the-art GRASP2K package.\cite{VRG:jonsson:2013:CPC}

\begin{table}[h]
    \centering
    \begin{tabular}{l|rrr}
    \hline\hline 
Basis$^a$ & \multicolumn{1}{c}{LCAO} & \multicolumn{1}{c}{MRA} & \multicolumn{1}{c}{FD} \\
\hline
2s1p &	-14.556 089 &	-14.616 856 &	-14.616 854 \\
3s2p1d &	-14.617 410 &	-14.654 415 &	-14.654 406 \\
4s3p2d1f &	-14.623 808 &	-14.661 475 &	-14.661 478 \\ \hline
complete & \multicolumn{3}{c}{-14.667 354 7$^b$} \\
\hline\hline
    \end{tabular}

    $^a$ Gaussian computations used the standard cc-pV\{D,T,Q\}Z\cite{VRG:prascher:2011:TCA} basis sets, respectively. MRA orbitals were obtained by optimization starting from the Gaussian AO basis sets.
    \\
    $^b$ Ref. \citenum{VRG:sims:2011:PRA}.
    \caption{CI energies ($mE_{\rm h}$) of the ground-state Be atom evaluated with fixed pre-optimized Gaussian AOs (``LCAO'') and the matching numerical orbitals optimized using the approach described in this work (``MRA'') and the standard atomic finite-difference MCSCF solver in the GRASP2K software (``FD'').\cite{VRG:jonsson:2013:CPC}}
    \label{tab:be-energies}
\end{table}

\subsection{Molecules}

\subsubsection{Bond breaking in HF and H$_2$O}

Valence CASSCF computations on the standard small-molecule benchmarks (hydrogen fluoride, water) were performed to illustrate the robustness of the real-space solver for computation on general (nonlinear) molecules with varying character of electron correlation.
Namely, CASSCF energy were determined with bond distances stretched from its equilibrium value ($R_e=\{0.9168,0.9572\} \AA$ for HF and H$_2$O, respectively) to twice the equilirbium value; the 104.52 degree bond angle in the latter was kept constant.

\begin{figure}[!ht]
\begin{tabular}{c}
  \includegraphics[width=0.7\textwidth]{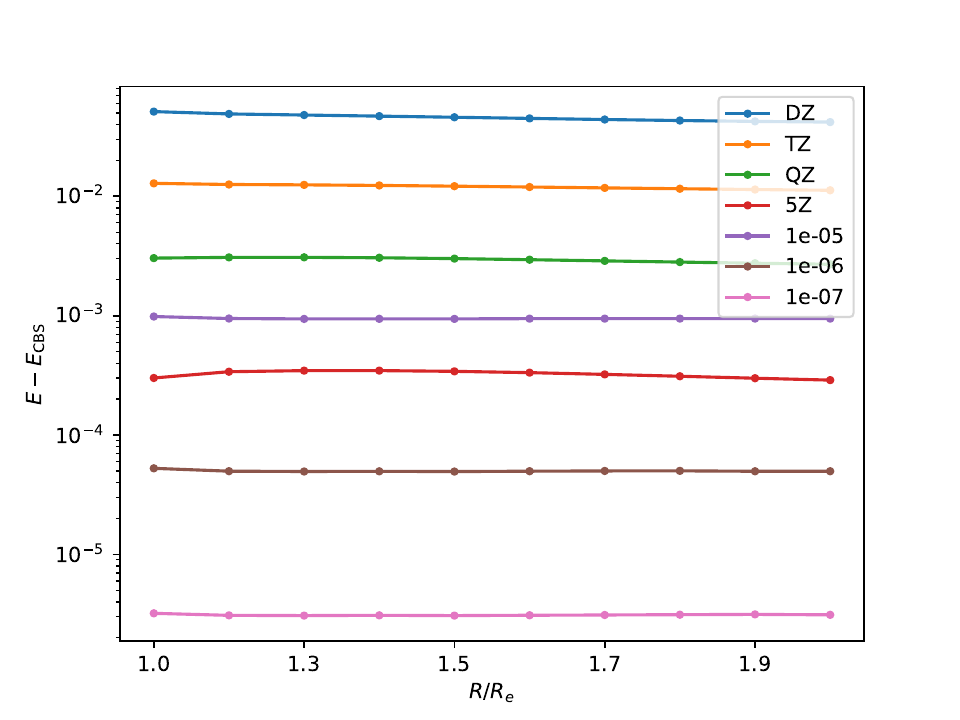} \\
  (a) HF\\
  \includegraphics[width=0.7\textwidth]{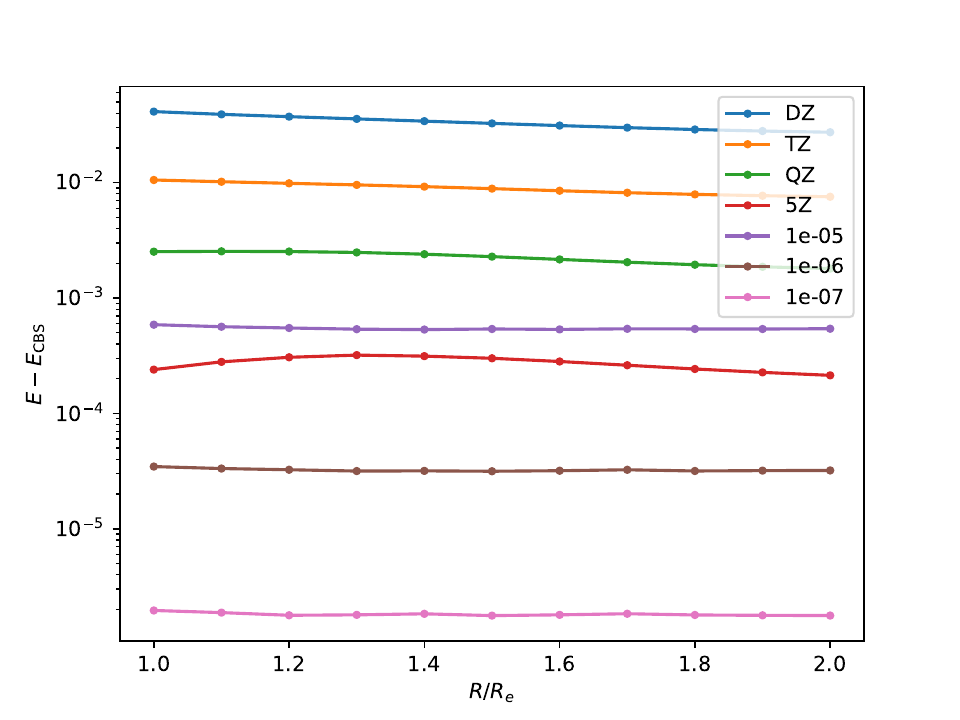} \\
  (b) H$_2$O
\end{tabular}
\caption{Basis set errors ($E_h$) of valence CASSCF energies obtained with Gaussian AO (cc-pV$X$Z bases\cite{VRG:dunning:1989:JCP}) and real-space orbital representations with varying precision parameter $\epsilon$ for single (a) and double (b) bond stretching. Complete basis set values were estimated by the real-space valence CASSCF energies obtained with $\epsilon=10^{-8}$.}
\label{fig:bond-breaking-cas}
\end{figure}

\Cref{fig:bond-breaking-cas} illustrates the basis set errors as a function of the bond distance for Gaussian AO CASSCF energies (obtained using the open-source \code{BAGEL} package\cite{VRG:shiozaki:2018:WCMS} with the cc-pV$X$Z-JKFIT density fitting bases used to approximate the 2-electron integrals) as well as our MRA representation with varying precision $\epsilon$. Already for $\epsilon=10^{-5}$ the real-space representation is more accurate than the cc-pVQZ Gaussian basis. Further tightening of $\epsilon$ reduced the error below that of cc-pV5Z Gaussian result, with the most precise energies obtained with the MRA approach, $\epsilon=10^{-8}$, likely converged to below a microhartree.

The logarithmic scale needed to illustrate the vast range of basis set errors ($>20 mE_\mathrm{h}$ with the cc-pVDZ basis, few $\mu E_\mathrm{h}$ with MRA $\epsilon=10^{-7}$) obscures the strong (and sometimes nonmonotonic) variation of basis set errors of the Gaussian AO representations. For example, the nonparallelity errors (defined as the largest minus the smallest error in the range) of the cc-pVDZ CASSCF energies are \{9.4, 13.9\} $mE_\mathrm{h}$ for HF and H$_2$O , respectively, and are large fractions of the corresponding \{51.3, 41.2\} $mE_\mathrm{h}$ maximum basis set errors. Importantly, the MRA energies have significantly smaller ratios of largest basis set errors to the nonparallelity errors than the Gaussian AO counterparts. For $\epsilon=10^{-5}$ the nonparallelity errors are only \{45, 54\} $\mu E_\mathrm{h}$ respecitively, compared to the \{984, 587\} $\mu E_\mathrm{h}$ largest basis set errors in the range; the corresponding values for the cc-pV5Z CASSCF are \{58, 106\} and \{347, 320\} $\mu E_\mathrm{h}$, respectively. The errors in MRA energies seem to cancel more systematically than the Gaussian AO counterparts, despite the intuition suggesting the latter should exhibit superior cancellation of errors.

\subsubsection{H$_{10}$}\label{sec:results_h10}
We chose H$_{10}$ in its ground state as a model system for testing solver robustness due to the availability of extensive benchmarks for a number of methods\cite{VRG:motta:2017:PRX}. The system offers the ability to control the nature and strength of electron correlation by varying the interatomic distance and the size of the orbital basis. In particular, at short interatomic distances ($R$) even the smallest nonminimal (double-zeta)  Gaussian AO basis becomes nearly linearly dependent and accurate evaluation of the Hamiltonian and the method-specific iterative solvers can become untenable. At short $R$ the importance of angular correlation greatly increases compared to the equilibrium and stretched $R$. Lastly, the use of a minimal AO basis (1 orbital per atom) maximizes the degree of ``on-site'' repulsion and increases the extent of longitudinal correlation.

The initial investigation of our orbital solver focused on the maximally-compressed geometry considered in Ref. \citenum{VRG:motta:2017:PRX}, where achieving the basis set limit with the standard Gaussian basis sets is difficult due to the rapid onset of ill-conditioning, which recently motivated developments of mixed numerical representations marrying Gaussian AOs with real-space approaches.\cite{VRG:stoudenmire:2017:PRL} As expected, the orbital optimization greatly reduces the basis set incompleteness of the CI energy, as can be seen in the rightmost column of  \cref{tab:h10-energy}, by $\times 3.4$ for the STO-6G basis an by $\times 7.4$ for the cc-pVDZ basis. The estimated residual basis set error with the MRA-optimized cc-pVDZ orbitals is $<3 mE_{\rm h}$ / atom; this is already comparable to the $~1 mE_{\rm h}$ uncertainty in the estimated nonrelativistic energy at this distance.\cite{VRG:motta:2017:PRX}

\begin{table}[hbt]
    \centering
    \begin{tabular}{l|rr|rrrrr}
\hline \hline
Basis$^a$ & \multicolumn{2}{c|}{HF} & \multicolumn{5}{c}{CI} \\
 & \multicolumn{2}{c|}{LCAO}  & \multicolumn{2}{c}{LCAO$^b$} & \multicolumn{2}{c}{MRA} & $\Delta_{\rm LCAO} / \Delta_{\rm MRA}$ \\
\hline
1s & -3,751.7	& (406.3) & -3,824.4 & (604.0) & -4,251.5 & (176.9) & 3.4 \\
2s1p& -4,010.1	& (147.9) & -4,219.7& (208.9) & -4,404.0 & (24.4) &  8.6 \\
3s2p1d & -4,139.3	& (18.7) & -4,394.5 & (33.9) &	-4,421.4 & (7.0) & 4.8 \\
4s3p2d1f& -4,150.5 & (7.5) & --- & &	--- & &  \\ \hline
complete&	\multicolumn{2}{c|}{-4,158.0} & \multicolumn{5}{c}{-4,428.4$^c$} \\
\hline\hline
    \end{tabular}

    $^a$ Gaussian computations used the standard STO-3G\cite{VRG:hehre:1969:JCP} and cc-pV\{D,T,Q\}Z\cite{VRG:dunning:1989:JCP} basis sets. The MRA orbitals were obtained by optimization starting from the corresponding Gaussian AO basis sets.
    \\
    $^b$ The corresponding MRCI+Q values from Ref.  \citenum{VRG:motta:2017:PRX} are -3,824.4 (STO-6G) and -4,219.5 (cc-pVDZ); MRCI did not converge with triple- and quadruple-zeta bases. \\
    $^c$ Estimated CBS auxiliary field quantum Monte Carlo (AFQMC) energy from Ref. \citenum{VRG:motta:2017:PRX}.
    \caption{HF and CI energies ($mE_{\rm h}$) of the equidistant 10-atom chain of H atoms at  compressed geometry (interatomic separation of 1 a.u.; see Ref. \citenum{VRG:motta:2017:PRX} for more details) evaluated with Gaussian AOs (``LCAO'') and optimized real-space orbitals (``MRA''; $\epsilon=10^{-5}$, bootstrapped from the corresponding Gaussian AO basis), as well as the ratio of the respective basis set errors.}
    \label{tab:h10-energy}
\end{table}

\begin{figure}[!ht]
\begin{tabular}{c}
  \includegraphics[width=\textwidth]{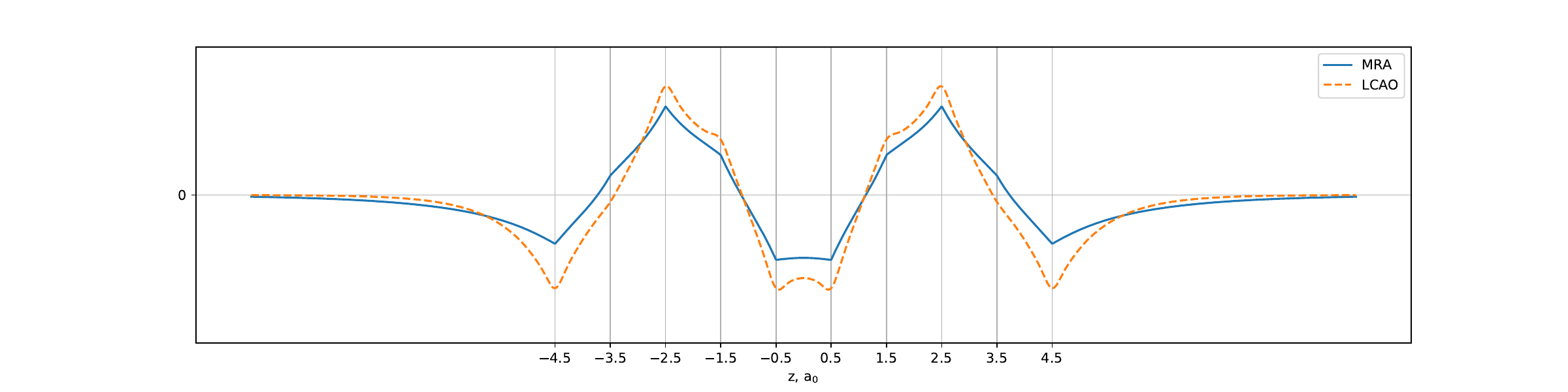} \\
  (a) 5th NO\\
  \includegraphics[width=\textwidth]{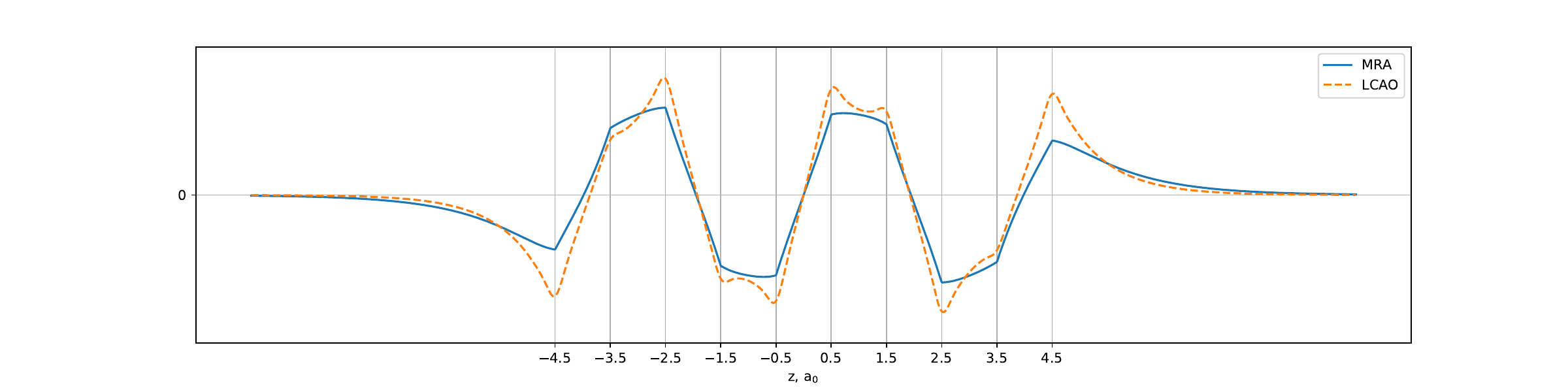} \\
  (b) 6th NO
\end{tabular}
\caption{Plots of frontier NOs of H$_{10}$ evaluated with STO-6G Gaussian AOs (``LCAO'') and the corresponding optimized real-space orbitals (``MRA'') along the molecular axis.}
\label{fig:h10-orbitals}
\end{figure}

\cref{fig:h10-orbitals} contains the plots of the frontier natural orbitals (counterparts of the HOMO and LUMO Hartree-Fock orbitals). The qualitative structures of the LCAO and real-space NOs are similar: both exhibit the same number of nodes and the same symmetry with respect to the inversion center. As expected, and just like in the atoms, the real-space NOs exhibit correct nuclear cusps and slower asymptotic decay into the vacuum. Interestingly, however, real-space NOs have lower probability densities at most nuclei than their Gaussian counterparts.

\section{Summary and Perspective}\label{sec:summary}

We presented a general method for constructing orbitals that are optimal for computing energies with correlated (many-body methods) in arbitrary molecules and with robust control of numerical errors. In this work we employed a multiresolution adaptive discontinuous spectral-element representation for the orbitals, but the presented method can be used with other real-space representations. Although utilized with the conventional and selected configuration interaction wave functions, the orbital solver can be used with arbitrary maps from orbitals to 1- and 2-particle reduced density matrices (e.g., tensor networks, stochastic methods, or quantum circuits). Numerical performance of the method was demonstrated by computing the CI energies of small atoms (matching the numerical precision of the atomic finite-difference CI codes) and a paradigmatic compressed polyatomic molecule; in all cases the use of real-space representation for the orbitals yielded energies that were significantly lower than the Gaussian AO-based counterparts, with correct asymptotic behavior at the nuclei and in vacuum.

Despite using the efficient production-grade implementation of the MRA calculus in \code{MADNESS}, the current implementation is not fully optimal. Although as we demonstrated here optimization of more than 100 {\em correlated} orbitals is already feasible with the current (CPU-only) code on a single multicore node, the implementation of the time-determining step in the orbital update, namely, the cumulant-dependent contribution to \cref{eq:phi-update} (evaluated at the $\mathcal{O}(n_\mathrm{act}^4)$ asymptotic cost, with $n_\mathrm{act}$ the number of active orbitals), does not explicitly optimize for CPU caches and can be significantly improved (by one to two orders of magnitude) even without exploiting accelerators. The optimization of $~140$ correlated orbitals in H$_{10}$ reported in \cref{sec:results_h10} took less than 4 hours per iteration on a single node of the \code{TinkerCliffs} cluster at Virginia Tech's Advanced Research Computing equipped with a 128-core AMD EPYC 7702 processor; this was already comparable with $~30$ minutes per iteration that the selected CI 2-RDM evaluation took.

The implementation can also be improved in the other asymptotic regime, namely for applications with a modest number of active orbitals ($n_\mathrm{a} \leq 50$) and a large number of inactive  orbitals, $n_\mathrm{i}$. The asymptotic cost is then dominated by the evaluation of the exchange contribution to the $\hat{v}\ket{\phi_m}$ term on the right-hand side of \cref{eq:phi-update}. By orbital localization of the inactive orbitals fast application of the exchange operator will be possible, as demonstrated by the past development of the optimally-scaling MRA Hartree-Fock method\cite{VRG:yanai:2004:JCPa}. In the current implementation the inactive orbitals are not localized, hence the cost of the exchange operator application is at worst $\mathcal{O}(n^3_\mathrm{c})$. Nevertheless practical applications to general systems can be readily performed without these optimizations. For example, a (4,4)-CASSCF($\epsilon=10^{-5})$ evaluation of the ground state energy of the butadiene molecule took 380 seconds on an Apple MacBook Pro with a 12-core Apple M2 Max chip; for comparison, the state-of-the-art Hartree-Fock MRA solver in \code{MADNESS} took 130 seconds. We expect the described improvements to bring down the cost of the CAS computation with small active spaces close to the cost of the optimized MRA Hartree-Fock implementation.
The conventional (4,4) CASSCF wave function of comparable quality requires a cc-pV5Z basis (see \cref{tab:c4h6}); such computation with \code{BAGEL}\cite{VRG:shiozaki:2018:WCMS} (using the matching cc-pV5Z-JKFIT density fitting basis) on the same machine took 186 seconds.

\begin{table}[hbt]
    \centering
    \begin{tabular}{l|r}
\hline \hline
Basis$^a$ & $\delta E^b$ \\ \hline
DZ & 54.05\\
TZ & 12.57\\
QZ & 2.73\\
5Z & 0.38\\
MRA($\epsilon=10^{-5}$) & 0.73 \\
MRA($\epsilon=10^{-6}$) & 0.06 \\
\hline\hline
    \end{tabular}

    $^a$ $X$Z denotes the cc-pV$X$Z Gaussian AO basis set\cite{VRG:dunning:1989:JCP}.
    \\
    $^b$ Error in $mE_\mathrm{h}$ relative to the MRA($\epsilon=10^{-7}$) energy, $-155.041338$ $E_\mathrm{h}$.
    \\
    \caption{Basis set errors of the ground state (4,4) CASSCF energies evaluated in LCAO and MRA representations for the (1,3)-butadiene molecule at the experimentally-derived equilibrium geometry.\cite{VRG:haugen:1966:ACS}}
    \label{tab:c4h6}
\end{table}

After the outlined algorithmic improvements we expect the cost of orbital optimization to become largely trivial relative to the wave function/density optimization. We are thus hopeful that the real-space representations will become competitive with or even preferred to the AO representations for correlated models.

Our formalism should map straightforwardly to the optimization of molecular spinors for relativistic correlated methods by borrowing the advances in 4-component relativistic numerical electronic structure introduced by some of us recently.\cite{VRG:anderson:2019:JCP} Extension to periodic systems is also straigtforward by using the existing extension of Poisson and Helmholtz Green's operators to periodic boundary conditions.\cite{VRG:jia:2009:P8ISDCABES} Work along these lines is already underway and will be reported elsewhere.

\begin{acknowledgement}
This work was supported by the U.S. Department of Energy via award DE-SC0022327. We are grateful to Greg Beylkin for providing the quadratures described in Ref. \citenum{VRG:ahrens:2009:PRSA}.
The authors acknowledge Advanced Research Computing at Virginia Tech ({\tt https://arc.vt.edu/}) for providing computational resources and technical support that have contributed to the results reported within this paper.
\end{acknowledgement}

\begin{appendices}

\section{Notation and Standard Definitions}
\label{sec:notation}
In this work we largely follow a covariant tensor notation of the many-body quantum chemistry.\cite{VRG:harris:1981:PRA,VRG:kutzelnigg:1982:JCP} Einstein summation convention is implied, unless sums over any index are shown explicitly: namely, summation is implied over every symbol that appears once in a {\em contravariant} (upper, bra) position and once as in a {\em covariant} (lower, ket) position in a given tensor {\em product}, with the summation range defined by the symbol. Spin-orbital basis is assumed throughout the paper for simplicity (the actual implementation is spin-adapted).

2- and 1-RDMs for state $\ket{\Psi}$ are defined as:
\begin{align}
    \gamma^{ij}_{kl} \equiv & \bra{\Psi} \hat{a}^i \hat{a}^j \hat{a}_l \hat{a}_k \ket{\Psi}, \\
    \gamma^i_j \equiv & \bra{\Psi}\hat{a}^i \hat{a}_j\ket{\Psi} \equiv \frac{2}{N-1}\sum_k \gamma_{ik}^{jk}, \label{eq:1rdm}
\end{align}
with $\hat{a}^i \equiv \hat{a}^\dagger_i$ and $\hat{a}_i$ the standard creators and annihilators, respectively.
If $\ket{\Psi}$ is a multideterminantal/multiconfiguration (MC) state it is convenient to represent
2-RDM as
\begin{align}
\label{eq:def-cumulant2}
    \gamma^{ij}_{kl} \equiv \, \gamma^i_k \gamma^j_l - \gamma^i_l \gamma^j_k + \lambda^{ij}_{kl},
\end{align}
with $\lambda^{ij}_{kl}$ the 2-RDM {\em cumulant}.\cite{VRG:kutzelnigg:1999:JCP}

Matrix elements of one-, two-, and higher-body operators are denoted in tensor notation as follows:
\begin{align}
\bra{p} \hat{o}(1) \ket{q} \equiv & \int \phi_p^*(1) \hat{o}(1) \phi_q(1) d1 \equiv o^q_p \\
\bra{p_1 p_2} \hat{o}(1,2) \ket{q_1 q_2} \equiv & \int \phi_{p_1}^*(1) \phi_{p_2}^*(2) \hat{o}(1,2) \phi_{q_1}(1) \phi_{q_2}(2) d1 d2 \equiv o^{q_1 q_2}_{p_1 p_2}, \text{etc.}
\end{align}
It is also convenient to use the tensor notation for potentials generated by orbital products:
\begin{align}
\hat{o}_{p_2}^{q_2} \ket{\phi_{q_1}} \equiv
\left( \int \phi_{p_2}^*(2) \hat{o}(1,2) \phi_{q_2}(2) d2 \right) \phi_{q_1}(1) .\label{eq:hatop2idx}
\end{align}
All many-body operators in this manuscript are particle-symmetric, i.e. $o(1,2) = o(2,1)$, hence $o^{p_1 p_2}_{q_1 q_2} = o^{p_2 p_1}_{q_2 q_1}$.
Matrix elements of the core (1-particle) Hamiltonian $\hat{o}(1) \equiv \hat{h}(1)$, composed of the free-particle Hamiltonian $\hat{d}$ and the Coulomb potential due to the nuclei $\hat{v}_n$,
are denoted by $h^p_q$. The corresponding matrix elements of the Coulomb 2-body interaction, $o(1,2) \equiv |\mathbf{r}_1 - \mathbf{r}_2|^{-1}$, are denoted by $g^{pq}_{rs}$. $s^p_q$ denotes the orbital overlap $\braket{\phi_p}{\phi_q}$. $\delta^p_q$ denotes the Kronecker delta:
\begin{align}
    \delta^p_q = & \begin{cases} 1, \quad p=q \\
    0, \quad p\neq q
    \end{cases}
\end{align}

The Fock operator has the usual definition:
\begin{align}
\label{eq:f}
\hat{f} \equiv \hat{h} + \hat{j} - \hat{k}.
\end{align}
$\hat{h} \equiv \hat{d} + \hat{v}_n$ is the 1-particle (core) Hamiltonian, composed of the free particle Hamiltonian $\hat{d}$ (i.e., the kinetic energy in the nonrelativistic case, or the Dirac Hamiltonian in the relativistic case) and the Coulomb potential due to the nuclei $\hat{v}_n$. $\hat{j}$, and $\hat{k}$ in \cref{eq:f} the Coulomb and exchange contributions, respectively:
\begin{align}
\hat{j}  \equiv & V \times, \\
\label{eq:k}
\hat{k} \ket{\phi_p} \equiv & \gamma_n^m \hat{g}^p_m \ket{\phi_n},
\end{align}
with $V$ the Coulomb potential generated by the electron density $\rho(\mathbf{r})$. Key to the approach in this paper is the partitioning of the Fock operator into the free particle Hamiltonian $\hat{d}$ and the potential $\hat{v} \equiv \hat{v}_n + \hat{j} - \hat{k}$:
\begin{align}
    \label{eq:f-dv}
    \hat{f} = \hat{d} + \hat{v}.
\end{align}
Evaluation of the electron density from the orbitals,
\begin{align}
    \rho(\mathbf{r}) \equiv \braket{\mathbf{r}}{\phi_m} \gamma_m^n \braket{\phi_n}{\mathbf{r}},
\end{align}
as well as evaluation of the Coulomb operator (\cref{eq:k}) is efficient in the {\em natural} basis which makes the 1-RDM diagonal.
The matrix elements of the Fock operator thus take the familiar form:
\begin{align}
    f^p_q \equiv \, h^p_q + \left(g^{pr}_{qs} - g^{pr}_{sq} \right) \gamma^s_r.
\end{align}

\section{Spherical projection of orbital updates}\label{sec:spherical-projection}

In atomic computations it is useful to be able to obtain solutions with specific target number of subshells, e.g. 3s2p1d for a second-row atom. In general such solutions are saddle points; tracking such saddle points can fail if the gradient does not preserve the relevant symmetries exactly. Unfortunately the finite-precision spectral element bases employed in this work do not preserve spherical symmetry. Thus we need to impose symmetry restrictions on the orbital updates; without such restrictions the solutions may break spherical symmetry.

Consider a subshell of $2l+1$ ``noisy'' orbitals. 
To project out the components of the orbitals corresponding to other angular momenta we want to express each atomic orbital as a product of its radial component $f(r)$ times the angular factor, i.e. as
\begin{align}
\phi_i ({\bf r}) \approx f(r) \sum_m^{2l+1} u_{i}^m Y_{lm}(\theta,\phi),
\end{align}
where the angular component is written as a unitary linear combination of spherical harmonics to support arbitrary orientation of each orbital.
We determine $f(r)$ 
by minimizing the average deviation of the subshell orbitals:
\begin{align}
    \Delta \equiv \sum_i^{2l+1} ||\phi_i ({\bf r}) - f(r) \sum_m^{2l+1} u_{i}^m Y_{lm}(\theta,\phi)||_2.
\end{align}
The gradient with respect to $f$ is:
\begin{align}
    \frac{\delta \Delta}{\delta f} = & 2 \sum_i^{2l+1} \braket{\phi_i - f \sum_m^{2l+1} u_{i}^m Y_{lm}}{\sum_{m'}^{2l+1} u_{i}^{m'} Y_{lm'}} \nonumber \\
    = & 2  \sum_{m'}^{2l+1} u_{i}^{m'} \sum_i^{2l+1} \braket{\phi_i}{Y_{lm'}} - 2 f \sum_i^{2l+1} \sum_m^{2l+1} u^{i}_m \sum_{m'}^{2l+1} u_{i}^{m'} \delta^m_{m'} \nonumber \\
    = & 2  \sum_{m'}^{2l+1} u_{i}^{m'} \sum_i^{2l+1} \braket{\phi_i}{Y_{lm'}} - (4l+2) f
\end{align}
This suggests the following
formula for $f$ for the given fixed coefficients $u$:
\begin{align}
    f = & \frac{1}{2l+1} \sum_{m'}^{2l+1} u_{i}^{m'} \sum_i^{2l+1} \braket{\phi_i}{Y_{lm'}}.
\end{align}
To determine nearly-optimal
values of coefficients $u$ for a given subshell orbital $i$ we evaluate its overlaps with the spherical harmonics $S^{(i)}_{m,g} \equiv \braket{\phi_i}{Y_{lm}}(r_g)$ on a uniform radial grid $\{r_g = g \delta_r, \forall g=1..n_g\}$. Here we set the grid size $n_g$ to 10 and the grid extent $\delta_g (n_g+1)$ to $\sqrt{\frac{2 \pi^2}{T_i}}$, which is the de Broglie wavelength of a free particle with kinetic energy $T_i \equiv \bra{\phi_i}\hat{T}\ket{\phi_i}$. Note that these kinetic energies are available since detection of orbital subshells also used their kinetic energies. Coefficients $u_i^{m'}$ are given by the left-hand singular vector corresponding to the largest singular value of $({\bf S}^{(i)})_{m,g} \equiv S_{m,g}$. Angular integration is performed via the lowest-order (72 points) rotationally invariant quadratures of Ahrens and Beylkin\cite{VRG:ahrens:2009:PRSA}.

\end{appendices}

\bibliography{vrgrefs}

\newpage
\thispagestyle{empty}
\includegraphics[width=\textwidth]{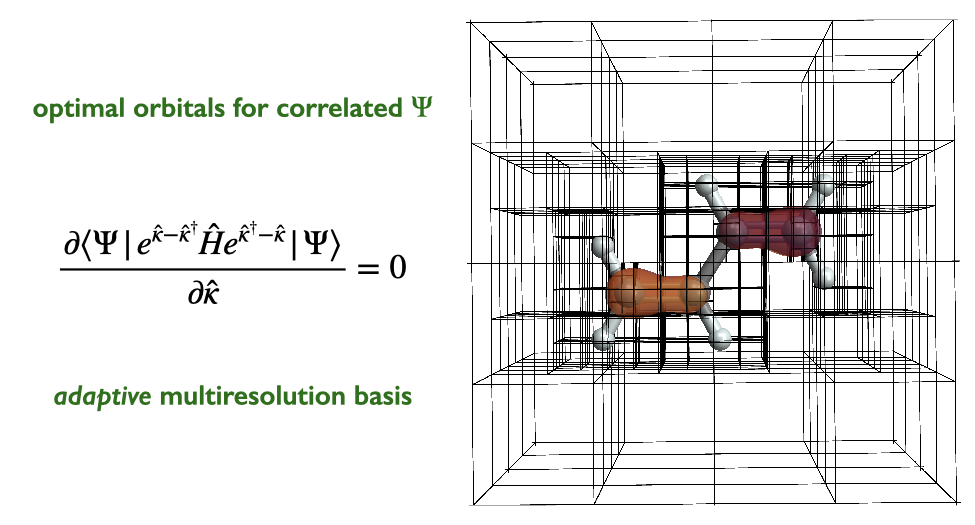}

\end{document}